\RequirePackage{ifpdf}
\documentclass[12pt,a4paper,hyper]{JHEP3}
\usepackage{epsfig}
\usepackage{amsmath,amsfonts,amssymb}
\usepackage{multirow}
\usepackage[titletoc]{appendix}
\usepackage{slashed}
\usepackage{graphicx}
\usepackage{epsfig}
\usepackage{psfrag}
\usepackage{color}

\usepackage{ulem}
\RequirePackage{filecontents} 

\def\={\; = \;}

\font\manual=manfnt
\def\dbend{\lower3.5pt\hbox{\manual\char127}}

\def\bar{\overline}

\def\CN{{\cal N}}
\def\aM{\bar{\text{M}}}
\def\M{\text{M}}

\title{\center{Generalized Kloosterman Sums from M2-branes }}

\preprint{}
\author{ Jo\~ao Gomes\\
	
\it Institute of Physics, University of Amsterdam,
Science Park 904, Postbus 94485, 1090 GL Amsterdam, The Netherlands \\ \vskip .5mm
\it Institute for Theoretical Physics, University of Utrecht, Princetonplein 3584 CC Utrecht, The Netherlands\\ \vskip .5mm

Email:
\email{J.M.VieiraGomes at uva.nl}
}

\abstract{Kloosterman sums play a special role in analytic number theory, for expressing the integer Fourier coefficients of modular forms as an infinite sum of Bessel functions, also known as Rademacher formula. The generalization to vector-valued modular forms is known as generalized Kloosterman sums. In the paper arxiv:1404.0033, a remarkable connection between these arithmetic sums and quantum black hole entropy was found. Nevertheless, the computation was particular for one-eighth BPS black holes in $\mathcal{N}=8$ string theory, which have a simple counting formula. Here, we review this construction and extend it to the case of one-quarter BPS black holes in $\CN=4$ string theory, which are counted by (mock) Jacobi forms of arbitrary index. The main result is an holographic derivation of the Kloosterman sums which includes the intricate sum over phases, and depends exactly on the spectral flow sectors and the spectrum of polar states. On the microscopic side we derive an analytic formula for the Kloosterman sums valid for any index, whereas from the macroscopic side we reproduce the same formula from the M-theory path integral on $\mathbb{Z}_c$ orbifolds of $AdS_2\times S^1$. A key aspect of the derivation is the identification of the spectral flow sectors with the contribution of M2 branes wrapping cycles on the compactification manifold.  After a careful treatment of the measure, the sum over orbifolds results in the sum over Bessels, in perfect agreement with the Rademacher expansion at any order in  perturbation theory.  }

\keywords{holography, supergravity, Localization}

\begin{document}
\section{Introduction}

Kloosterman sums are arithmetic sums of the form
\begin{equation}\label{sec1 Kloosterman}
 Kl(n,m,c)=\sum_{\substack{d\in (\mathbb{Z}/c\mathbb{Z})^* \\ ad=1\text{ mod}(c)}}\exp{\left[2\pi i\frac{d}{c} n+2\pi i \frac{a}{c} m\right]},
\end{equation}
 for integers $n,m,c$. These sums appeared originally in the problem of representing large numbers in quadratic forms of four variables \cite{Kloos}.  However, they occur most notably in the  Hardy-Ramanujan-Rademacher expansion \cite{Rademacher-1938,Zuckerman-Rademacher}, which we review later.

Recently, Kloosterman sums were shown to be related to non-perturbative corrections to the Bekenstein-Hawking area formula of BPS black holes \cite{Dabholkar:2014ema}. Following previous work on quantum black hole entropy \cite{Sen:2008vm} and localization techniques in supergravity \cite{Dabholkar:2010uh,Dabholkar:2011ec}, the authors of \cite{Dabholkar:2014ema} were able to identify the sums (\ref{sec1 Kloosterman}) with additional saddles in the path integral, related to global contributions on $AdS_2\times S^1/\mathbb{Z}_c$ orbifolds. In particular, the exponential  factor and the different sums in (\ref{sec1 Kloosterman}) were shown to arise after a careful evaluation of the Chern-Simons action of the flat connections living on the orbifold. Topologically, the orbifold corresponds to a Dhen filled solid torus parametrized by the integers $c,d$, and so the sum over those integers in (\ref{sec1 Kloosterman}) can be understood as a sum over topologies in quantum gravity. 

The essential step in uncovering the Kloosterman sums, is the application of the localization technique in the string theory path integral that computes the quantum entropy \cite{Dabholkar:2010uh,Gupta:2012cy}. This allows for an exact computation of the black hole entropy as function of the charges. This way we have control over the non-perturbative corrections, which is where the Kloosterman sums become more relevant.

Modular forms with non-positive weight have the remarkable property that its Fourier coefficients can be written as an infinite sum of Bessel functions, each of which comes multiplied by Kloosterman sums. This is known as Rademacher expansion \cite{Rademacher-1938}, and its generalization to vector-valued modular forms is the generalized Rademacher expansion \cite{Dijkgraaf:2000fq,Manschot:2007ha}. The idea behind the generalized version consists in writing the Jacobi form as a sum over theta functions $\theta_{\mu}(\tau,z)$ multiplied by vector-valued modular forms $f_{\mu}(\tau)$ \cite{Eichler:1985ja}. Then, the application of the circle method to the vector-valued modular forms gives the generalized Rademacher expansion. The Kloosterman coefficients (\ref{sec1 Kloosterman}) are modified to account for the fact that vector-valued modular forms transform among themselves under modular transformations. The generalized sums are schematically of the form
 \begin{equation}\label{sec1 Generalized Kloos}
 Kl(n,m,c)_{\mu\nu}=\sum_{\substack{d\in (\mathbb{Z}/c\mathbb{Z})^* \\ ad=1\text{mod} c}}e^{2\pi i (n-\frac{\mu^2}{4k})\frac{d}{c}}M^{-1}(\gamma)_{\nu\mu}e^{2\pi i (m-\frac{\nu^2}{4k})\frac{a}{c}},\;\gamma=\left(\begin{array}{cc}a & b\\
                         c & d                                                                                                                                                                                                                                                                                                                                                                  \end{array}\right)\in SL(2,\mathbb{Z}),
\end{equation}
where $\mu,\nu$ is an index in the space of vector-valued modular forms, $M(\gamma)_{\mu\nu}$ is the matrix that maps vector-valued modular forms to themselves under the modular transformation, and $k$ is the index of the Jacobi form.  In physical terms, we can identify the Jacobi form with the elliptic genus of the underlying microscopic CFT. The index $k$ is usually of the order of the central charge and $\mu,\nu$ parameterize spectral flow sectors.  

The main goal of this work is to provide with a bulk string theory computation of the generalized Kloosterman sums (\ref{sec1 Generalized Kloos}), for arbitrary index $k$. This extends the results of \cite{Dabholkar:2014ema} to one-quarter BPS black holes in $\CN=4$ string theory. We shall have in mind though that the black hole partition function in the $\CN=4$ theory is a mock Jacobi form \cite{Dabholkar:2012nd}, and so the usual Rademacher expansion does not apply. Nevertheless for the range of charges we will be considering the exact answer is well approximated by a Jacobi form \cite{Ferrari:2017msn}\footnote{We thank Atish Dabholkar for clarifying this point.}. In any case, the Kloosterman sums depend only on general transformation properties of the (mock)-Jacobi forms, and not on particular details. This is the feature that we want to reproduce from the bulk theory.

The analysis of \cite{Dabholkar:2014ema} focused on the case of one-eighth BPS black holes, which have a simple counting formula \cite{Maldacena:1999bp}. In this case, the black hole degeneracies  are the Fourier coefficients of the weak Jacobi form 
\begin{equation}\label{sec1 Jacobi form}
\phi_{-2,1}(\tau,z)=\frac{\vartheta(\tau,z)^2}{\eta^{6}(\tau)},
\end{equation} 
where $\vartheta(\tau,z)$ is a theta function and $\eta(\tau)$ is the Dedekind function. $\phi_{-2,1}(\tau,z)$ is a weak Jacobi form of weight minus two and index one. The generalized Kloosterman sums (\ref{sec1 Generalized Kloos}), in particular the matrix $M(\gamma)_{\mu\nu}$ can be constructed starting with the representation of $M$ for the generating elements $S,T$ of $SL(2,\mathbb{Z})$, and then building a general expression from the decomposition $\gamma=ST^{n_1}ST^{n_2}\ldots\,\in SL(2,\mathbb{Z})$. This was done in \cite{Dabholkar:2014ema} with the help of a result by Jeffrey \cite{Lisa92} in the context of compact Chern-Simons theory and Witten invariants \cite{Witten:1988hf}. However, the computation was specific to index one Jacobi forms such as (\ref{sec1 Jacobi form}). For arbitrary index, a similar computation is possible but it is much more challenging. Part of our work is devoted to obtaining an analytic formula for the matrix $M(\gamma)$ valid for any index, which we can use to compare with the bulk computation. Our formula is based on a result developed long time ago by  H. D. Kloosterman \cite{10.2307/1969082}, which we use extensively. In the appendix we provide with an independent proof of that formula. 

From the gravity point of view, the different sums  and phases in (\ref{sec1 Generalized Kloos}), for the index one Jacobi form (\ref{sec1 Jacobi form}), can be shown to arise from the contribution of flat connections to a Chern-Simons action on the Dhen filled solid torus $\simeq AdS_2\times S^1/\mathbb{Z}_c$ \cite{Dabholkar:2014ema}. This Chern-Simons action contains both "gravitational" \footnote{This means the usual map from three dimensional gravity and $SL(2)$ Chern-Simons theory.} and gauge Chern-Simons terms.
However,  the Chern-Simons level, which maps to the index of the Jacobi form, is an arbitrary charge dependent parameter, whereas the index of the microscopic counting function is one. So to obtain agreement between the bulk computation and the microscopic prediction (\ref{sec1 Jacobi form}), one has to fix the Chern-Simons level to be exactly one \cite{Dabholkar:2014ema}. This is puzzling in view of the U-duality invariance of the microscopic formula. On the other hand, we need very large central charge, and thus large Chern-Simons level, for the theory to have a semiclassical description. Our work will provide with the steppingstones to tackle this puzzle completely.

To achieve the main goal of the paper, we develop on the proposal \cite{Gomes:2017eac} for computing the exact quantum entropy of one-quarter BPS black holes. Essentially, the proposal provides a bulk physical interpretation for the non-perturbative corrections to the entropy, related to the polar coefficients of the vector-valued modular forms. It is argued that the path integral receives, besides the attractor geometry, additional saddles, which result from quantum fluctuations of the Calabi-Yau manifold. These fluctuations  lead, in turn, to a renormalization of the parameters that define the effective five dimensional Lagrangian, from which we compute the path integral using localization. Furthermore, the geometry gets corrected in such a way that only a finite number of geometries contribute. The bound on this number is also the bound imposed by the stringy exclusion principle \cite{Maldacena:1998bw}. The great advantage of this construction is that we can identify each of the Bessel functions, associated with the polar terms in the Rademacher expansion, with the perturbative quantum fluctuations around each new saddle. Then it becomes natural from the path integral point of view to include also orbifolds of those geometries, which is what we do in this work. 

Following \cite{Gomes:2017eac}, the fluctuations of the Calabi-Yau can be described equivalently  in terms of M2 and anti-M2 branes  wrapping cycles on the Calabi-Yau and sitting at the origin of $AdS_2\times S^1$; this picture is borrowed from the chiral primary counting of \cite{Gaiotto:2006ns,Gaiotto:2006wm}. It is found that the difference between the number of M2 and anti-M2 ($\aM2$) branes generates a large gauge transformation on the $U(1)$ gauge fields of supergravity. However, such gauge transformations are singular on the disk $\simeq AdS_2$.  As a consequence, the holonomies around the contractible cycle, that is, the around the disk, change to account for the presence of the M2 branes; when the same number of M2 and $\aM2$ is present, the total charge is zero and the gauge transformation vanishes. Following closely \cite{Dabholkar:2014ema}, we use this description to compute the contribution of these holonomies to the Chern-Simons action. As a result, one obtains precisely the generalized Kloosterman sums for arbitrary index $k$. In particular, we identify the spectral flow sectors $\nu$ in (\ref{sec1 Generalized Kloos}) with the singular gauge transformations.  

In addition, we provide a simple derivation of the localization measure to include the effect of the orbifold geometries. This generalizes the result of \cite{Gomes:2015xcf} to the case of $AdS_2\times S^1/\mathbb{Z}_c$ orbifolds, and we use this to fix the dependence of the localization finite dimensional integral on the  order of the orbifold $|\mathbb{Z}_c|=c$. Such dependence is crucial for the convergence of the full answer, for the following reason. Note that, in a large charge expansion of the black hole degeneracy $d(q,p)$, the orbifold saddles lead to corrections of the form
\begin{equation}
\sim \exp{\left[\frac{A}{4c}+\ldots\right]},\;\;A/c\gg 1
\end{equation}
where $A$ is the area of the horizon and the $\ldots$ denote perturbative corrections around each saddle orbifold geometry. Clearly, for sufficiently large $c$ and fixed $A$, the saddle point approximation breaks down. However, using localization one can show that such contributions are of order one, and so the sum over these order one terms leads to a potential divergence, unless the measure is correctly taken into account. It is important to stress that this divergence can not be studied using perturbative methods for the reason just explained, and only a non-perturbative off-shell computation such as localization can provide such test.

Putting together the contribution coming from the localization computation, that is, the Bessel functions, the generalized Kloosterman sums and the $|\mathbb{Z}_c|$ dependent measure, we obtain precisely the Rademacher expansion. 

The plan of the paper is as follows.  In section \S \ref{sec2 Gen Kloos}, we review the generalized Rademacher expansion and associated generalized Kloosterman sums. The main result is an analytic formula for the multiplier matrix, which is the core of the generalized sums. Then in section \S \ref{sec3 Holo computation}, we describe the holographic computation using an effective three dimensional Chern-Simons theory. A crucial step in this exercise is the inclusion of the singular gauge transformations that signal the presence of the M2 and $\aM2$ branes. We show that this leads precisely to the structure of the Kloosterman sums. Finally in section  \S \ref{sec Measure}, we derive the dependence of the measure on the order of the $\mathbb{Z}_c$ orbifold. We show this agrees precisely with the Rademacher expansion.

\section{Generalized Kloosterman sums}\label{sec2 Gen Kloos}

In this section, we review the generalized Rademacher expansion for the Fourier coefficients of vector-valued modular forms \cite{Dijkgraaf:2000fq,Manschot:2007ha}. Later we provide with an analytic formula for the generalized Kloosterman sums.

Recently, an extension of the Rademacher expansion to mock-Jacobi forms was considered \cite{Ferrari:2017msn}. The structure of this expansion is very similar to the usual Rademacher expansion of Jacobi forms, in the sense that we have a sum over Bessel functions dressed by Kloosterman sums. However, in the mock case, the sum contains additional Bessel functions, whose index differs from the Bessels that appear in the usual Rademacher expansion (\ref{sec2 generalized Rademacher}). In particular, these Bessels have integral index for integer weight $\omega$. For our purpose, we will only be considering the Bessels of half-integer index, which are common to both Jacobi and mock-Jacobi forms. In both the mock and Jacobi examples, the Kloosterman sums are determined by general modular transformation properties, and so, for our purpose, it is enough to consider the Jacobi case.  

\subsection{Generalized Rademacher expansion}

 A Jacobi form $\varphi(\tau,z)$ of weight $\omega$ and index $k$ satisfies the transformation properties
\begin{equation}\label{modular transf}
 \varphi\left(\frac{a\tau+b}{c\tau+d},\frac{z}{c\tau+d}\right)=(c\tau+d)^{\omega}e^{2\pi i k\frac{cz^2}{c\tau+d}}\varphi(\tau,z),\;\;\left(\begin{array}{cc}a & b\\
                         c & d                                                                                                                                                                                                                                                                                                                                                                  \end{array}\right)\in SL(2,\mathbb{Z}),
\end{equation}
and
\begin{equation}\label{sec2 elliptic sym}
 \varphi(\tau,z+l\tau+m)=e^{-2\pi i k(l^2\tau+2lz)}\varphi(\tau,z),\;\;\;l,m\in \mathbb{Z},
\end{equation}
also known as elliptic symmetry. Using the property (\ref{sec2 elliptic sym}) we can decompose the Jacobi form as a sum over theta functions \cite{Eichler:1985ja}, that is,
\begin{equation}\label{sec2 theta decomposition}
 \varphi(\tau,z)=\sum_{\mu\,\text{mod }2k}h_{\mu}(\tau)\theta_{\mu,k}(\tau,z),
\end{equation}
where the theta functions are defined as
\begin{equation}\label{sec2 theta fnct}
 \theta_{\mu,k}(\tau,z)=\sum_{n\in\mathbb{Z}}q^{k\left(n+\mu/(2k)\right)^2}y^{\mu+2kn},\;\;\;q=e^{2\pi i \tau},\,y=e^{2\pi i z},
\end{equation}
The functions $h_{\mu}(\tau)$ are vector-valued modular forms and have the Fourier expansion
\begin{equation}\label{Fourier vector modular}
 h_{\mu}(\tau)=q^{-\Delta_{\mu}}\sum_{n=0}^{\infty}H_{\mu}(n)q^{n}.
\end{equation}
The part of $h_{\mu}(\tau)$ with negative $n-\Delta_{\mu}$ is called the polar part.

Under modular transformations the theta functions transform to themselves in the following way
\begin{equation}\label{theta modular transf}
 \theta_{\mu,k}\left(\frac{a\tau+b}{c\tau+d},\frac{z}{c\tau+d}\right)=(c\tau+d)^{1/2}e^{2\pi i k\frac{cz^2}{c\tau+d}}\sum_{\nu\text{ mod }2k}M^{-1}(\gamma)_{\mu\nu}\theta_{\nu,k}(\tau,z).
\end{equation}
The matrix $M(\gamma)_{\mu\nu}$ is called the multiplier system. For the generating elements $S,T\in SL(2,\mathbb{Z})$, one has respectively
\begin{eqnarray}
 &&\theta_{\mu,k}(-1/\tau,z/\tau)=\sqrt{\frac{\tau}{2k i}}e^{2\pi i k\frac{z^2}{\tau}}\sum_{\nu\,\text{mod }2k}e^{-\pi i \frac{\mu\nu}{k}}\theta_{\nu,k}(\tau,z),\\
&&\theta_{\mu,k}(\tau+1,z)=e^{\pi i\frac{\mu^2}{2k}}\theta_{\mu,k}(\tau,z).
\end{eqnarray}
Given the modular transformation property of the Jacobi form (\ref{modular transf}) together with (\ref{sec2 theta decomposition})  and (\ref{theta modular transf}), one finds that the functions $h_{\mu}(\tau)$ transform as
\begin{equation}
 h_{\mu}\left(\frac{a\tau+b}{c\tau+d}\right)=(c\tau+d)^{\omega-1/2}\sum_{\nu\text{ mod }2k}M(\gamma)_{\nu\mu}h_{\nu}(\tau),
\end{equation}
which justifies the name vector-valued modular form. We see that the multiplier matrix $M(\gamma)_{\mu\nu}$ is a representation of $SL(2,\mathbb{Z})$ in the space of vector-valued modular forms. 

The generalized Rademacher expansion is an exact formula for the Fourier coefficients of $\varphi(\tau,z)$. 
 Given the theta function decomposition, it is easy to show that the Fourier coefficients are given in fact by $H_{\mu}(n)$ (\ref{Fourier vector modular}).  Following \cite{Dijkgraaf:2000fq,Manschot:2007ha}, we have 
\begin{eqnarray}\label{sec2 generalized Rademacher}
 H_{\mu}(n)=&&\frac{1}{i^{\omega+1/2}}\sum_{m-\Delta_{\nu}<0}H_{\nu}(m)\sum_{c=1}^{\infty}\frac{1}{c} Kl(n,m,c)_{\mu\nu}\\
&&\times \int_{\epsilon-i\infty}^{\epsilon+i\infty}\frac{dt}{t^{5/2-\omega}}\exp{\left[2\pi\frac{(n-\Delta_\mu)}{c t}-2\pi \frac{(m-\Delta_{\nu})}{c}t\right]},
\end{eqnarray}
where $m-\Delta_{\nu}<0$ defines the polarity and $H_{\nu}(m)$ is the associated polar coefficient. The function $Kl(n,m,c)_{\mu\nu}$ are the generalized Kloosterman sums 
\begin{equation}\label{sec2 Gen Kloosterman}
 Kl(n,m,c)_{\mu\nu}=\sum_{\substack{0\leq -d<c;\,(d,c)=1\\ ad=1\text{ mod}(c)}}e^{2\pi i (n-\Delta_{\mu})\frac{d}{c}}M^{-1}(\gamma)_{\nu\mu}e^{2\pi i (m-\Delta_{\nu})\frac{a}{c}},
\end{equation}
with no implicit sum on $\mu,\nu$.

In the case of modular forms, the generalized Kloosterman sum (\ref{sec2 Gen Kloosterman}) reduces to the classical definition (\ref{sec1 Kloosterman}). To see this, suppose we have a modular form with Fourier expansion
\begin{equation}
 f(\tau)=q^{-n_p}\sum_{n\geq 0}^{\infty}d(n)q^{n},
\end{equation}
with $n_p>0$. In this case, $n-\Delta_{\mu}$ and $m-\Delta_{\nu}<0$ in (\ref{sec2 Gen Kloosterman}) are replaced respectively by $n>0$ and $m-n_p<0$, which is the polarity. Moreover, since we are dealing with a modular form,  we do not have spectral flow sectors $\mu,\nu$, and hence there is no multiplier matrix. Therefore, the Kloosterman sum reduces to
\begin{equation}
 Kl(n,m,c)=\sum_{\substack{0\leq -d<c;\,(d,c)=1\\ ad=1\text{ mod} (c)}}e^{2\pi i n\frac{d}{c}+2\pi i (m-n_p)\frac{a}{c}}.
\end{equation}

\subsection{Analytic formula for the Multiplier Matrix}

To construct an analytic formula for the matrix $M(\gamma)_{\mu\nu}$ we can build a general representation starting with the generating elements of $SL(2,\mathbb{Z})$. For the case with index $k=1$ this was done in  \cite{Dabholkar:2014ema} following a result by Jeffrey \cite{Lisa92} in the context of Chern-Simons theory. However for general index $k$ the problem is technically more challenging and we cannot straightforwardly use the results of \cite{Lisa92}. Fortunately, this problem was solved long time ago by H.D. Kloosterman \cite{10.2307/1969082}, which provides an explicit representation for the matrix $M^{-1}(\gamma)_{\mu\nu}$. This is very convenient because $M^{-1}(\gamma)_{\mu\nu}$ appears explicitly in the Kloosterman sums (\ref{sec2 Gen Kloosterman}). In the appendix \S\ref{appendix A} we give an alternative derivation of that formula.

In \cite{10.2307/1969082}, H. D. Kloosterman provides many results on the transformation of generalized theta functions under modular transformations. For our purpose, we are interested in equations 2.15, 3.5 and 3.8 of that paper. Specializing his results to the  theta functions of index $k$  (\ref{sec2 theta fnct}), one obtains the expression
\begin{equation}\label{sec2 Kloos rep}
 M^{-1}\left(\gamma\right)_{\mu\nu}=\frac{1}{(2kc i)^{1/2}}\sum_{m=0}^{c-1}\exp{\left[2\pi i\left(\frac{a}{c}\frac{(\mu+2k m)^2}{4k}-\frac{\nu (\mu+2k m)}{2kc}+\frac{d}{c}\frac{\nu^2}{4k}\right)\right]},
\end{equation}
with $\gamma=\left(\begin{array}{cc}
a & b\\
c & d 
 \end{array}\right) \in SL(2,\mathbb{Z})$. This is one of the main formulas of our work.

The representation (\ref{sec2 Kloos rep}) has a few important properties that will be useful later on. First one has 
\begin{equation}\label{prop 1}
M^{-1}(\gamma)_{\mu+2k l,\nu}=M^{-1}(\gamma)_{\mu\nu},\;\;l\in \mathbb{Z},
\end{equation}
and similarly
\begin{equation}\label{prop 2}
M^{-1}(\gamma)_{\mu,\nu+2kl}=M^{-1}(\gamma)_{\mu\nu},\;\;l\in \mathbb{Z},
\end{equation}
which ia the statement that the representation (\ref{sec2 Kloos rep}) only depends on the equivalence class of $\mu,\nu\in \mathbb{Z}/2k\mathbb{Z}$. Second we have 
\begin{equation}
 \sum_{\sigma=0}^{2k-1}M^{-1}(\gamma)_{\mu\sigma}M^{-1}(\gamma')_{\sigma\nu}=M^{-1}(\gamma\gamma')_{\mu\nu},\;\;\gamma,\gamma'\in SL(2,\mathbb{Z}).
\end{equation}
This follows from the fact that (\ref{sec2 Kloos rep}) is a representation of $SL(2,\mathbb{Z})$. In the appendix we show explicitly how the representation (\ref{sec2 Kloos rep}) obeys this property. We also give derivations of the properties (\ref{prop 1}) and (\ref{prop 2}).

\section{Holographic computation}\label{sec3 Holo computation}

In this section we describe the holographic dual computation of the generalized Kloosterman sums using Chern-Simons theory on $AdS_2\times S^1/\mathbb{Z}_c$ orbifolds. The discussion is very similar to \cite{Dabholkar:2014ema}, which we briefly review  now.

The $AdS_2\times S^1/\mathbb{Z}_c$ orbifolds were studied originally in \cite{Murthy:2009dq} by considering a decoupling limit of the $SL(2,\mathbb{Z})$ family of extremal black hole solutions in $AdS_3$ \cite{Maldacena:1998bw}. The inclusion of the orbifold geometry in the path integral explains non-perturbative corrections to black hole entropy of the form
\begin{equation}\label{orbifold entropy}
\sim \exp{\left[\frac{A}{4c}\right]},
\end{equation}
with $A$ the horizon area; the factor of $1/c$ is a direct consequence of the orbifold. The orbifold consists in identifying points on $AdS_2$ which differ by a deficit of $2\pi/c$ angle, while performing a translation along the circle $S^1$ by $2\pi d/c$, with $d,c$ coprime; the translation along the circle renders the quotient smooth. Globally one has a solid torus $D\times S^1$, with $D$ a disk, filled with a hyperbolic metric. A choice of $(c,d)$ is equivalent to choose which cycle in the boundary torus we are making contractible in the full geometry. That is, after choosing a basis of one-cycles $C_1$ and $C_2$ on the boundary torus, we Dhen fill the solid torus by attaching a disk to a cycle $C_c$, which becomes the contractible cycle; this is a linear combination of the basis one-cycles,
\begin{equation}
C_c\equiv cC_1+dC_2,
\end{equation}
whereas the non-contractible circle $S^1$ is identified with the linear combination
\begin{equation}
C_{nc}\equiv aC_1+bC_2.
\end{equation}
To guarantee that $C_{nc}$ has unit intersection with $C_c$, that is, $C_{nc}\cap C_c=1$ given $C_1\cap C_2=1$, we must have $ad-bc=1$, with $a,b,c,d \in\mathbb{Z}$ . From now on we denote the orbifold geometry by $M_{(c,d)}$.

In \cite{Dabholkar:2014ema}, it is shown that the quantum entropy path integral receives the contribution of flat connections on $M_{(c,d)}$ via their Chern-Simons action. The non-trivial feature of the computation is that, while the local contributions to the path integral give rise to contributions to the entropy that are real and of the form (\ref{orbifold entropy}), the flat connections, on the other hand, give rise to phases, essentially  because the Chern-Simons action is not parity invariant. In particular, the  Chern-Simons action of the flat connections in the $M_{(c,d)}$ geometry gives rise to the phases that one finds in the Kloosterman sums of (\ref{sec1 Jacobi form}) \cite{Dabholkar:2014ema}. For each $c$, we have to sum over $d$ and $a$, which is the element inverse of $d$ in $\mathbb{Z}/c\mathbb{Z}$. This explains the various sums in the Kloosterman formula.

Since the contribution of the flat connections is topological in nature, it is enough to consider the effective Chern-Simons action living on the solid torus defined by the geometry $M_{(c,d)}$. Moreover, we can show that the Chern-Simons action depends only on the holonomies of the flat connection along the contractible and non-contractible cycles, which simplifies greatly the discussion.

Following \cite{Dabholkar:2014ema}, we consider the $SL(2,\mathbb{R})_L\times SL(2,\mathbb{R})_R\times SU(2)_L\times SU(2)_R$ effective Chern-Simons action living on the geometry $M_{(c,d)}$. The theory contains in addition multiple $U(1)$ Chern-Simons terms but they do not contribute to the entropy because the action of an abelian flat connection is zero. The non-compact  $SL(2,\mathbb{R})_L\times SL(2,\mathbb{R})_R$ gauge group factor comes from the fact that three dimensional gravity can be written in terms of Chern-Simons variables with the gauge group being determined by the isometries of $AdS_3$. Supersymmetry acts on the right, that is, on the $SL(2,\mathbb{R})_R\times SU(2)_R$ factor, with $SU(2)_R$ the R-symmetry. Furthermore, one has an $SU(2)_L$ factor, which arises from gauging the isometries of a local $S^3$ in the full geometry. 

The Chern-Simons action contains the following terms
\begin{equation}
S=-\frac{i\tilde{k}_L}{4\pi}I[\tilde{A}_L]+\frac{i\tilde{k}_R}{4\pi}I[\tilde{A}_R]-\frac{ik_R}{4\pi}I[A_R]+\frac{ik_L}{4\pi}I[A_L],
\end{equation}
where $\tilde{A}_{L,R}$ are respectively the $SL(2,\mathbb{R})_{L,R}$ connections and $A_{L,R}$ are the $SU(2)_{L,R}$ connections. We weight the path integral with $\exp S$. Due to supersymmetry the Chern-Simons levels $\tilde{k}_R$ and $k_R$ are equal. Nevertheless, the levels $\tilde{k}_L$ and $k_L$ remain independent. We have denoted the Chern-Simons action by $I[A]$, which we define as
\begin{equation}
I[A]=\int_M \text{Tr}\left(A\wedge dA+\frac{2}{3}A^3\right),
\end{equation}
with the trace in the fundamental representation.

To compute the Chern-Simons action on the solid torus we follow \cite{kirk1990,Dabholkar:2014ema}. A flat connection $A_f$ is always pure gauge and as such we can write it as
\begin{equation}
A_f=-dg g^{-1},\;g\in \mathbf{G},
\end{equation}
where $\mathbf{G}=SL(2),SU(2)$ is the gauge group. As explained in \cite{kirk1990} the gauge transformation $g$ can be brought to the "normal" form
\begin{equation}
g=f(x_c,r)e^{-\frac{i}{2}\beta \sigma^3 x_{nc}}.
\end{equation}
The coordinates $(x_c,r)$ with $x_c\in [0,2\pi]$ and $r\in[0,1]$, parametrize the disk $D$, while $x_{nc}$ parametrizes the circle $S^1$. The function $f(x_c,r)\in \mathbf{G}$ maps points on the disk to elements of the gauge group. One has the condition that $f(x_c,r=1)=e^{-\frac{i}{2}\alpha \sigma^3 x_c}$ at the boundary of the disk, and moreover it is constant at the origin, so that the gauge field is well defined at that point. The constant $\beta$ fixes the holonomy along the non-contractible cycle. With this parametrization the holonomies become diagonal in $\mathbf{G}$, up to conjugation.

The Chern-Simons action of the flat connection can be computed using the Stoke's theorem \cite{kirk1990}. This gives
\begin{equation}
I[A_f]=2\pi^2\alpha\beta.
\end{equation}
Furthermore, in the path integral with Chern-Simons action we need to introduce the boundary action \cite{Elitzur:1989nr,Hansen:2006wu,Dabholkar:2014ema} 
\begin{equation}\label{CS bnd term}
S_{\text{bnd}}=\frac{1}{4\pi}\int_{\partial M}\text{Tr}A_1A_2,
\end{equation}  
where $A_1$ is the component of $A$ along the cycle $C_1$ in the boundary torus and similarly for $A_2$. This boundary action ensures that the variational problem is well posed. These boundary values are determined as follows. In the black hole problem, the leading contribution to the entropy comes from the geometry $M_{1,0}$. In this case, the cycle $C_1$ bounds the $AdS_2$ disk and it is parametrized by the euclidean time, whereas, the M-theory circle corresponds to the cycle $C_2$, and is parametrized by the coordinate $y$. The boundary conditions are such that the component along $C_2$ is fixed, while the component along $C_1$, that is, $A_1$ is allowed to fluctuate. In \cite{Dabholkar:2014ema}, this choice has been shown  to be consistent with the $AdS_2$ microcanonical boundary conditions of the quantum entropy formalism \cite{Sen:2008vm}. 

For the manifold $M_{(c,d)}$, $\alpha$ and $\beta$, which parametrize respectively the holonomies along the contractible and non-contractible cycles, are determined as in \cite{Dabholkar:2014ema}, that is
\begin{equation}\label{Cc cycle}
2\pi i\frac{\sigma^3}{2}\alpha\equiv\oint_{C_c}A_f=c\oint_{C_1}A_f+d\oint_{C_2}A_f,
\end{equation}
and
\begin{equation}\label{Cnc cycle}
2\pi i\frac{\sigma^3}{2}\beta\equiv\oint_{C_{nc}}A_f=a\oint_{C_1}A_f+b\oint_{C_2}A_f,
\end{equation}
computed at $\partial{M}=T^2$. Define 
\begin{equation}
\oint_{C_1}A_f=2\pi i\gamma \frac{\sigma}{2},\; \oint_{C_2}A_f=2\pi i\delta \frac{\sigma}{2},
\end{equation}
then from equations (\ref{Cc cycle}) and (\ref{Cnc cycle}) we obtain
\begin{equation}\label{holonomies alpha beta}
\alpha=c\gamma+d\delta,\;\beta=a\gamma+b\delta.
\end{equation}
The Chern-Simons action of the flat connection together with the boundary term (\ref{CS bnd term}) is
\begin{equation}\label{CS action plus bnd}
S+S_{\text{bnd}}=\frac{i\pi}{2}k\alpha\beta-\frac{i\pi}{2}k\gamma\delta,
\end{equation}
where we have reintroduced the Chern-Simons level $k$.

\subsection{Flat connections from M2 branes}

 In \cite{Gomes:2017eac}, it was proposed that the quantum entropy path integral of M-theory on $AdS_2\times S^1\times S^2\times M_6$, with $M_6$ a Calabi-Yau manifold, receives the contribution of a finite number of off-shell backgrounds. The contribution of these saddles to the path integral can be computed using localization, and it turns out, that they can be identified with the polar Bessel functions of the Rademacher expansion. It is argued that after turning on singular fluxes on the Calabi-Yau, the full back-reacted geometry is a solution of five dimensional supergravity with renormalized $c_2$ coefficient, which parameterizes the mixed gauge-gravitational Chern-Simons terms in five dimensions. When the fluxes are absent, $c_2$ is the second Chern-class (tangent bundle) of the Calabi-Yau. The presence of fluxes can be interpreted equivalently in terms of M2 and anti-M2 branes wrapping holomorphic cycles in the Calabi-Yau. The renormalization is such that the effective Chern-Simons levels are 
\begin{equation}\label{renormalized levels}
\tilde{k}_R=\frac{p^3}{6}+\frac{\hat{c}_2\cdot p}{12},\;\tilde{k}_L=\frac{p^3}{6}+\frac{\hat{c}_2\cdot p}{6},
\end{equation}
and similarly for $k_R$, which is equal to $\tilde{k}_R$ by supersymmetry. The parameter $\hat{c}_2$ denotes the effective renormalized value of $c_2$, which in terms of the fluxes $f_a$ and $\bar{f}_a$ is given by
\begin{equation}
\hat{c}_{2a}=c_{2a}-12(f_a+\bar{f}_a),\quad f_a,\bar{f}_a\in \mathbb{Z}^+.
\end{equation}
where the subscript $a$ parameterizes a basis of two cycles. In the M2 brane picture, $f_a$ and $\bar{f}_a$ are the number of M2 and anti-M2 branes respectively. The range of $f_a,\bar{f}_a$ is not arbitrary. The renormalization leads to a correction of the physical size of the geometry which puts a bound on $f_a,\bar{f}_a$. In \cite{Gomes:2017eac}, this bound was shown to be the same as the one imposed by the stringy exclusion principle.

In addition,  such fluxes induce a large gauge transformation on the $U(1)$ gauge fields of five dimensional supergravity, as
\begin{equation}
A^a\rightarrow A^a-2(f^a-\bar{f}^a)dx_c,
\end{equation}
where we defined $f_a=D_{ab}f^b$ with $D_{ab}=D_{abc}p^c$, and $D_{abc}$ is the Calabi-Yau intersection matrix; $p^a$ are the magnetic fluxes on the sphere and map to the configuration of $\M5$ branes wrapping a four cycle on the Calabi-Yau \cite{Maldacena:1997de}. Note that the gauge transformation is singular at the origin since it is proportional to $dx_c$, which is the disk angle, and $A^a$ vanishes there. Physically this singularity is expected because there are M2 branes sitting at the origin, with $f_a-\bar{f}_a$ the total charge.

For the $\CN=4$ theory, which is our primary interest, we have
\begin{equation}\label{gauge transf M2}
\Delta f^1=\Delta f^1=-\frac{p^1}{P^2}(f_1-\bar{f}_1),\;\;\Delta f^a=0,\,a\neq 1
\end{equation}
where we have defined $\Delta f^a\equiv (f^a-\bar{f}^a)$; we also have  $D_{abc}=D_{1ab}=D_{a1b}=D_{ab1}=C_{ab}$. We can show that  the $U(1)$ gauge field $A^1$  corresponds to a $U(1)$ truncation of the $SU(2)_L$ gauge field after dimensional reduction on the sphere \cite{Dabholkar:2014ema}. The precise map is $A_L=i\sigma^3 A^1/2$ \cite{Dabholkar:2014ema,Gomes:2015xcf}. Following \cite{Dabholkar:2014ema}, we compute the $SU(2)_L$ Chern-Simons contribution. In this work we consider $p^1=1$ for simplicity. It would be important to generalize these results for arbitrary $p^1$, though we believe the results will not suffer significant changes. The boundary conditions for $A_L$ can be determined from the attractor equations. We have
\begin{equation}
\oint_{C_2}A_L=2\pi \mathbf{i}\frac{Q.P}{P^2},
\end{equation}
where $Q.P=-q_1p^1+q_ap^a$ and $P^2=C_{ab}p^ap^b$ with $a=2\ldots n_v$, and $n_v$ the number of vectors. We have denoted $\mathbf{i}\equiv i\sigma^3$.  A key aspect of our construction when compared with the Kloosterman sum computation of \cite{Dabholkar:2014ema}, is that  the Wilson line along the contractible cycle receives the contribution of the M2 branes that are sitting at the origin. Essentially, the gauge transformation (\ref{gauge transf M2}) leads to the Wilson line
\begin{equation}
\pi\mathbf{i}\alpha =\oint_{C_c}A_L=\pi \mathbf{i}\left(2n+\frac{\nu}{k_L}\right),\;\nu=\epsilon k_L+f_1-\bar{f}_1,
\end{equation}
with $\epsilon=\pm 1$, $k_L=P^2/2$ and $n\in \mathbb{Z}$. We have introduced $\epsilon$ such that in the absence of M2 branes the holonomy is $-\mathbf{1}$; this equals the holonomy of the $SU(2)_R$ connection \cite{Gomes:2015xcf}, which is necessary to ensure that the geometry corresponds to the R sector of the dual CFT. Given $\alpha=c\gamma+d\delta$ and $\beta=a\gamma+b\delta$ (\ref{holonomies alpha beta}) we determine
\begin{equation}
\gamma=\frac{1}{c}\left(2n+\frac{\nu}{k_L}\right)-\frac{d}{c}\frac{Q.P}{k_L},
\end{equation}
and hence
\begin{equation}
\beta=\frac{a}{c}\left(2n+\frac{\nu}{k_L}\right)-\frac{ad}{c}\frac{Q.P}{k_L}+b\frac{Q.P}{k_L},
\end{equation}
with $ad-bc=1$. Therefore the total Chern-Simons action plus boundary terms (\ref{CS action plus bnd}) is
\begin{equation}\label{SU2 CS phases}
I_{\text{CS+Bnd}}=\frac{\pi i}{2k_L}\frac{a}{c}\left(\nu+2n k_L\right)^2-\frac{\pi i}{k_L c}Q.P(\nu+2k_L n)+\frac{\pi i}{2k_L}\frac{d}{c}(Q.P)^2+2\pi i \mathbb{Z}.
\end{equation}
It is easy to see that the exponential of $I_{\text{CS+Bnd}}$ is invariant under $n\rightarrow n+c\mathbb{Z}$ and so we have to truncate the sum of $n$ to lie in $\mathbb{Z}/c\mathbb{Z}$. Similarly we can show (see appendix \ref{appendix A}) that the exponential is invariant under $Q.P\rightarrow Q.P+2k_L\mathbb{Z}$, and so we can write $Q.P=\mu$ with $\mu\in \mathbb{Z}/2k_L\mathbb{Z}$.

Now we consider the gravitational $SL(2)_{L,R}$ and $SU(2)_R$ Chern-Simons terms. On the supersymmetric side the contributions coming from the $SL(2,\mathbb{R})_R$ and $SU(2)_R$ terms cancel each other as pointed out in \cite{Dabholkar:2014ema}. The holonomy of $A_R$ is such that the orbifold preserves the localization supercharge. Nevertheless, the gravitational $SL(2,\mathbb{R})_L$ contribution is non-trivial an equals  (equation 4.46 in \cite{Dabholkar:2014ema})
\begin{equation}\label{eq SL CS contrib}
\exp{\left(-\pi i\frac{\tilde{k}_L}{2}\frac{a}{c}+\pi i\frac{\tilde{k}_L}{2}\frac{d}{c}R^2\right)},
\end{equation}
where $R$ is the asymptotic value of the radius of the M-theory circle and $\tilde{k}_L$ is the renormalized $SL(2,\mathbb{R})_L$ Chern-Simons level (\ref{renormalized levels}). The radius $R$ can be computed for $\Delta f=0$, which gives
\begin{equation}
R^2=\frac{\Delta}{k_L\tilde{k}_L},
\end{equation} 
after solving the attractor equations, or equivalently, from extremizing the quantum entropy function. Here $\Delta=Q^2P^2-(Q.P)^2$ is the quartic invariant charge combination. Hence, (\ref{eq SL CS contrib}) becomes
\begin{equation}\label{tree level SL a/c}
\exp{\left(-\pi i\frac{\tilde{k}_L}{2}\frac{a}{c}+\pi i\frac{\Delta}{2k_L}\frac{d}{c}\right)}.
\end{equation}

The Rademacher expansion predicts, nonetheless, the phase 
\[-i\frac{\pi}{2}\left(\tilde{k}_L-2(\Delta f)^2\right)\frac{a}{c}+\pi i\frac{\Delta}{2k_L}\frac{d}{c}\]
 with the combination $\tilde{k}_L-2(\Delta f)^2$ being the polarity. We have defined $(\Delta f)^2=D^{ab}\Delta f_a\Delta f_b$. In the five dimensional theory \cite{Gomes:2017eac}, the term $(\Delta f)^2$ in the polarity arises indirectly from a delta function contribution of the $U(1)$ gauge field strength, as explained in \cite{Gomes:2017eac}. To be more precise, from the localization computation of \cite{Gomes:2017eac} we find the entropy function 
\begin{equation}
-2\pi \frac{\hat{q}_0}{R}+\frac{\pi}{2} R\tilde{k}_L-\frac{\pi}{4}RD_{ab}(\phi^a+q^a\phi^0)(\phi^b+q^b\phi^0)-2\pi i\frac{\phi^a}{\phi^0}\Delta f_a -2\pi iq_a\Delta f^a,
\end{equation}
with $\hat{q}_0=q_0-D_{ab}q^aq^b/2=-\Delta/4k_L$ and $R=2/\phi^0$. The terms proportional to $\Delta f_a$ arise from the delta function induced by the large gauge transformation. When $\Delta f_a=0$, we can identify, at the on-shell level, the term $-2\pi \frac{\hat{q}_0}{R}+\frac{\pi}{2} R\tilde{k}_L$ with the Chern-Simons action of the flat connection on $M_{(1,0)}$. Integrating out $\phi^a$, the gaussian induces a correction $(\Delta f)^2$ to $\tilde{k}_L$, and one effectively obtains  the entropy
\begin{equation}
-2\pi \frac{\hat{q}_0}{R}+\frac{\pi}{2} R(\tilde{k}_L-2(\Delta f)^2).
\end{equation}
Since in the effective three dimensional Chern-Simons theory one assumes that both the $SL(2,\mathbb{R})_L$ and $SU(2)_L$ factors are decoupled, the effective Chern-Simons level for the $SL(2,\mathbb{R})_L$ factor should be in fact $\tilde{k}_L-2(\Delta f)^2$. On the other hand, from the attractor equations we have now $R^2=\Delta/k_L(\tilde{k}_L-2(\Delta f)^2)$. Proceeding as before, now we find the phase
\begin{equation}\label{quantum SL a/c}
\exp{\left(-\pi i\frac{\tilde{k}_L-2(\Delta f)^2}{2}\frac{a}{c}+\pi i\frac{\Delta}{2k_L}\frac{d}{c}\right)}.
\end{equation}

We can identify $\tilde{k}_L-2(\Delta f)^2$ with the polarity $\Delta_{\nu}-m>0$ in the Rademacher expansion (\ref{sec2 generalized Rademacher}). That is, in terms of the fluxes that polarity has the form
\begin{eqnarray}
\frac{\tilde{k}_L-2(\Delta f)^2}{4}&=&\frac{(P^2/2-(f_1-\bar{f}_1))^2}{2P^2}-\bar{f}_1+n_p\nonumber\\
&=&\frac{\nu^2}{4k_L}-m
\end{eqnarray}
with $\nu=k_L-(f_1-\bar{f}_1)$ and $m=\bar{f}_1-n_p$, and $n_p=0,1$ for the $T^4,K3$ CHL orbifold compactifications respectively \cite{Gomes:2017eac}.
 
 Assembling the different pieces, the phase (\ref{quantum SL a/c}) then gives the term that multiplies $M^{-1}(\gamma)$ in (\ref{sec2 Gen Kloosterman}). The charge combination $\Delta=Q^2P^2-(Q.P)^2$ can always be written in form $4nk_L-\mu^2$ with $Q.P=\mu \text{ mod}(2k_L)$ and $n\in \mathbb{Z}$. So we identify $\Delta/4k_L$ with $n-\Delta_{\mu}$ in (\ref{sec2 Gen Kloosterman}). Similarly we have $\Delta_{\nu}=\nu^2/4k_L$ in the same expression. Furthermore, integration over the $\phi^a$ gives rise to a term proportional to $1/\sqrt{\text{det}D_{ab}}$, which for the $\CN=4$ compactifications is proportional to $1/\sqrt{k_L}$, after setting $p^1=1$. The exponential  of the $SU(2)_L$ Chern-Simons contribution (\ref{SU2 CS phases}), together with the factor $1/\sqrt{k_L}$, reproduces the analytic formula for the matrix $M^{-1}(\gamma)_{\nu\mu}$, except for a dependence on $c$ in the normalization factor, which we fix in the next section.

\subsection{Measure dependence on $|\mathbb{Z}_c|$}\label{sec Measure}

The measure of the finite dimensional integral that one obtains using localization can be fixed by one-loop computation in Chern-Simons theory. The original computation \cite{Gomes:2015xcf} focused on the  $AdS_2\times S^1$ geometry but we can easily generalize it for the $AdS_2\times S^1/\mathbb{Z}_c$ orbifolds. We also point the reader to the discussion in section 5.1 of \cite{Gomes:2017eac}.

The result for the partition function in the unorbifolded theory is the integral
\begin{equation}\label{measure}
\int dR\prod_{a=1}^{b+1} d\phi^a\, \frac{1}{R}\,\exp{[-2\pi \frac{\hat{q}_0}{R}+\frac{\pi}{2} R\tilde{k}_L-\frac{\pi}{4}RD_{ab}\phi^a \phi^b]}.
\end{equation}
We are neglecting a factor dependent on the physical size of the geometry, which does not play any role for what we want to say. The measure $1/R$ follows from a one-loop computation in Chern-Simons theory, which gives
\begin{equation}\label{Z CS 1-loop}
Z^{\text{CS}}_{\text{1-loop}}\propto \frac{R^{-b/2-1}}{\sqrt{\prod k_i}},
\end{equation}
where $k_i$ runs through the $SL(2,\mathbb{R})_L\times SU(2)_L\times U(1)^{b}$ Chern-Simons levels. The one-loop contribution (\ref{Z CS 1-loop}) comes entirely from the zero modes of the gauge fields whose measure in the path integral is determined using an ultra-locality argument. Essentially, one imposes an ultra-local measure of the form
\begin{equation}\label{ultralocality normalization}
\int D[A]\exp{[k\int \text{Tr }A\wedge \star A]}=1,
\end{equation}
for the non-abelian gauge fields, and similarly for the $U(1)$ gauge fields. This normalization defines the measure $D[A]$. In Chern-Simons theory we have to pick a metric to define this measure, and this is the reason why the integral over the zero modes gives factors of $R$ in (\ref{Z CS 1-loop}).

We can repeat the same logic but now for the orbifold $AdS_2\times S^1/\mathbb{Z}_c$. This will give a dependence of $D[A]$ on the order of the orbifold $|\mathbb{Z}_c|=c$. We have to remark, nevertheless, that for the $U(1)$ gauge fields the dependence of the measure $D[A]$ on $c$ is ambiguous. The reason is that from the Chern-Simons point of view, the $U(1)$ gauge fields are free fields and so we could have absorbed a $1/c$ dependence of $\int \text{Tr }A\wedge \star A$ in the normalization (\ref{ultralocality normalization}), in a rescaling of the gauge fields. In contrast, the non-abelian gauge fields are interacting fields and a rescaling leads effectively to a change in the interaction term. Therefore, we understand that for the $U(1)$ gauge fields, we have the choice to exclude from $D[A]$ the $c$ dependence that one obtains from the ultralocality argument. We will see this leads to the desired result. Nevertheless, it would be important to check this more explicitly.

In the unorbifolded theory, the normalization (\ref{ultralocality normalization}) gives a dependence of $(k\, R)^{1/2}$ in the measure for each non-abelian gauge group factor, while in the orbifold case we have $(k R/c)^{1/2}$, where the $1/c$ factor is the result of the quotient. For the remaining $U(1)$ factors we obtain  a $(k R)^{1/2}$ dependence with no $c$ factor, as argued. Repeating the one-loop computation in the orbifold geometry, which is a volume over the zero modes modes, we find 
\begin{equation}
Z^{\text{CS}}_{\text{1-loop}}|_{M_{(c,d)}}\propto c\frac{R^{-b/2-1}}{\sqrt{\prod  k_i}},
\end{equation}
where the $c$ factor comes from the $SL(2,\mathbb{R})_L\times SU(2)_L$ gauge fields. On the other hand, the entropy function for the orbifold geometry is 
\begin{equation}
-2\pi \frac{\hat{q}_0}{cR}+\frac{\pi}{2} \frac{R}{c}\tilde{k}_L-\frac{\pi}{4}\frac{R}{c}D_{ab}\phi^a\phi^b,
\end{equation}
where the factor of $1/c$ is due to the $\mathbb{Z}_c$ quotient. For purpose of computing the localization measure we have set $\Delta f=0$. Comparing the Chern-Simons computation with the one-loop correction that we obtain from extremizing the entropy function, we find that the measure in (\ref{measure}) acquires an additional factor of $c^{-b/2}$.

Nonetheless, this computation only takes into account the fluctuations around a particular flat connection in the geometry $M_{(c,d)}$. The path integral contains for fixed $c$ a sum over geometries $M_{(c,d)}$, and holonomies $\text{Hol}(A_L)$, which are parameterized respectively by the integers $d$ and $n$ valued in $\mathbb{Z}/c\mathbb{Z}$ (\ref{SU2 CS phases}). From the Chern-Simons point of view, the sum over geometries and holonomies is characterized by the sum over the Wilson lines $\oint_{C_1}\tilde{A}_L$ and $\oint_{C_1}A_L$ respectively, which are allowed to fluctuate. Since these holonomies are  $\mathbb{Z}_c$ valued, each sum over  $\oint_{C_1}\tilde{A}_L$ and $\oint_{C_1} A_L$ must be accompanied by a factor of $1/c$, which ensures that the volume of the gauge group is correctly factored out. This gives an additional $1/c^2$ factor.

The finite dimensional integral then has the form
\begin{equation}
\frac{1}{c^{b/2+2}}\int dR\prod_{a=1}^{b+1} d\phi^a\frac{1}{R}\exp{[-2\pi \frac{\hat{q}_0}{cR}+\frac{\pi}{2} \frac{R}{c}\tilde{k}_L-\frac{\pi}{4}\frac{R}{c}D_{ab}\phi^a\phi^b]}.
\end{equation}
Performing the various gaussian integrals, we obtain the Bessel answer
\begin{equation}
\frac{1}{c\sqrt{c}\sqrt{\text{det}(D_{ab})}}\int \frac{dR}{R^{3/2+b/2}}\exp{[-2\pi \frac{\hat{q}_0}{cR}+\frac{\pi}{2} \frac{R}{c}\tilde{k}_L]}.
\end{equation}
The term $\sqrt{\text{det}(D_{ab})}$ is proportional to $\sqrt{k_L}$.  The factor $1/\sqrt{ck_L}$ joins the $SU(2)_L$ Kloosterman sum (\ref{SU2 CS phases}) to give the normalization found in the the matrix $M^{-1}(\gamma)$ (\ref{sec2 Kloos rep}). The remaining $1/c$ factor can be identified with the  one that multiplies the Kloosterman sums in the Rademacher formula (\ref{sec2 generalized Rademacher}). We therefore obtain an exact matching with the microscopic formula. By construction, we now have guaranteed that for large $c$ this measure ensures that the sum over the $M_{(c,d)}$ geometries is not divergent.

Given this result we can also try to reproduce the measure that one obtains for modular forms. In this case the Kloosterman sums reduce to the classical case (\ref{sec1 Kloosterman}). From the bulk point of view we also do not expect $SU(2)_L$ Chern-Simons terms. Following the same reasoning, we find a single factor of $1/c$ from the zero mode volume, since now we only have a sum over the $\oint_{C_1} \tilde{A}_L$ holonomies parameterized by $d\in \mathbb{Z}/c\mathbb{Z}$. Again, this result agrees with the Rademacher prediction.

\section{Discussion and Conclusion}

In this work, we have considered the contribution of $AdS_2\times S^1\times S^2/\mathbb{Z}_c$ orbifolds to the quantum entropy path integral following the proposal \cite{Gomes:2017eac}. We have provided a generalization of the non-perturbative corrections studied in \cite{Dabholkar:2014ema} for one-quarter BPS black holes in four dimensional $\CN=4$ string theory. To this end the main results are:
\begin{itemize}
\item {\it{Generalized Kloosterman sums}}: we have derived generalized Kloosterman sums from the gravity point of view.  A key aspect of this construction is the contribution of non-trivial flat connections to the Chern-Simons action, which arise after considering M2  and $\aM2$-branes wrapping cycles on the Calabi-Yau \cite{Gomes:2017eac}. The result of the bulk computation is in perfect agreement with an analytic formula for the generalized Kloosterman sums of Jacobi forms of arbitrary index.

\item {\it{Integers from quantum gravity}}:  we derive the exact dependence of the localization measure on the order of the orbifold $|\mathbb{Z}_c|=c$.  Together with the Kloosterman sums and the Bessel functions that we obtain by supersymmetric localization, we can show that the $AdS_2$ path integral reproduces the Rademacher expansion at all orders in the charges. 
  
\end{itemize}

Our results constitute a very important piece of evidence in favor of the proposal put forward in \cite{Gomes:2017eac}. It is important to stress the key aspect of that proposal which is at the heart of the construction presented in this work: the fact that we can associate to each polar Bessel function a different saddle geometry. Then, the inclusion of the orbifold geometries becomes straightforward from the path integral point of view and the Chern-Simons computation follows as originally shown in \cite{Dabholkar:2014ema}.

As a byproduct of our results it would be important to understand if the Kloosterman sums obey special arithmetic properties that can explain the black hole degeneracy of more general charge configurations. From the microscopic side we already have a good understanding of the degeneracy of non-primitive dyons for both $\CN=8$ and $\CN=4$ compactifications \cite{Sen:2008sp,Banerjee:2008pu,Dabholkar:2008zy}. On the macroscopic side, there is partial understanding \cite{Sen:2009vz,Sen:2009gy} in terms of  $AdS_2\times S^2$ orbifolds in the quantum entropy. However, this holds only at the on-shell level, and so it would be important to extend such results to the quantum level, as we did in this work. If such arithmetic properties exist, one may be able to solve a puzzle related to U-duality invariance of the $\CN=8$ answer raised in the beginning of this work.

\subsection*{Acknowledgments}

We would like to thank Atish Dabholkar and Jan Manschot for discussions on related topics. This work is part of the Delta ITP consortium, a program of the Netherlands Organisation for Scientific Research (NWO) that is funded by the Dutch Ministry of Education, Culture and Science (OCW).

\begin{appendices}

\section{An elementary derivation of the multiplier matrix}\label{appendix A}

In this section we use a trick by Zagier and Skoruppa \cite{Zagier1989} to derive an analytic formula for the multiplier matrix.

We start with the definition of the theta functions
\begin{eqnarray}
 \theta_{m,\rho}(\tau,z)&=&\sum_{l=\rho\,\text{mod}(2m)}q^{l^2/4m}y^l\\
&=&\sum_{n\in\mathbb{Z}}q^{(\rho+2m n)^2/4m}y^{(\rho+2mn)},
\end{eqnarray}
with $\rho$ a representative of the equivalence class $\mathbb{Z}/2m\mathbb{Z}$. These are modular functions with weight $1/2$ and level $m$, that is, under modular transformations, one has
\begin{eqnarray}
 &&\theta_{m,\rho}\left(\frac{a\tau+b}{c\tau+d},\frac{z}{c\tau+d}\right)=(c\tau+d)^{1/2}e^{2\pi i m\frac{cz^2}{c\tau+d}}\sum_{\sigma\text{ mod }2m}K_{\rho\sigma}(\gamma)\theta_{m,\sigma}(\tau,z)\label{theta modular}\\
&&\theta_{m,\rho}(\tau,z+\lambda \tau+\mu)=e^{-2\pi im(\lambda^2\tau+2\lambda z)}\theta_{m,\rho}(\tau,z),\;(\lambda,\mu)\in \mathbb{Z}^2,
\end{eqnarray}
with $\gamma=\left(\begin{array}{cc}
a & b\\
c & d 
\end{array}\right) \in SL(2,\mathbb{Z})$. For convenience, define
\begin{equation}\label{theta bar}
 (\theta_{m,\rho}|\gamma)(\tau,z)\equiv (c\tau+d)^{-1/2}e^{-2\pi i m\frac{cz^2}{c\tau+d}}\theta_{m,\rho}\left(\frac{a\tau+b}{c\tau+d},\frac{z}{c\tau+d}\right),
\end{equation}
which by (\ref{theta modular}), is equivalent to
\begin{equation}\label{def Kloos}
 (\theta_{m,\rho}|\gamma)(\tau,z)=\sum_{\sigma\text{ mod }2m}K_{\rho\sigma}(\gamma)\theta_{m,\sigma}(\tau,z).
\end{equation}

By definition, we have the equality
\begin{equation}
 (\theta_{m,\rho}|\gamma)(\tau,z)=\sum_{s\in \mathbb{Z}}y^s\int_{0}^1e^{-2\pi i x s}(\theta_{m,\rho}|\gamma)(\tau,x)dx.
\end{equation}
Introducing the expression (\ref{theta bar}) in this integral and using the Fourier expansion of the theta function, we obtain
\begin{eqnarray}
 &&(\theta_{m,\rho}|\gamma)(\tau,z)=\sum_{s\in \mathbb{Z}}y^s(c\tau+d)^{-1/2}\int_{0}^1e^{-2\pi i x s}e^{-2\pi i m\frac{cx^2}{c\tau+d}}\theta_{m,\rho}\left(\frac{a\tau+b}{c\tau+d},\frac{x}{c\tau+d}\right)dx\nonumber\\ \label{eq theta rewrite}
&=&\sum_{s\in \mathbb{Z}}y^s(c\tau+d)^{-1/2}\sum_{r=\rho\,\text{mod}(2m)}\int_{0}^1\exp{\left[2\pi i\left( -x s - m\frac{cx^2}{c\tau+d}+\gamma(\tau)\frac{r^2}{4m}+\frac{x}{c\tau +d}r\right)\right]}dx,\nonumber\\
{}
\end{eqnarray}
with $\gamma(\tau)=a\tau+b/c\tau+d$. Using 
$
 \frac{a\tau+b}{c\tau+d}=\frac{a}{c}-\frac{1}{c(c\tau+d)}$
we write
\begin{equation}
 -x s - m\frac{cx^2}{c\tau+d}+\gamma(\tau)\frac{r^2}{4m}+\frac{x}{c\tau +d}r=\frac{a}{c}\frac{r^2}{4m}-\frac{sr}{2mc}+\frac{d}{c}\frac{s^2}{4m}+\tau\frac{s^2}{4m}-\frac{mc}{c\tau+d}\left(x-\frac{r}{2mc}+s\frac{c\tau+d}{2mc}\right)^2\nonumber,
\end{equation}
and hence equation (\ref{eq theta rewrite}) becomes
\begin{eqnarray}
 (\theta_{m,\rho}|\gamma)(\tau,z)&=&\sum_{s\in \mathbb{Z}}\sum_{r=\rho\,\text{mod}(2m)} q^{s^2/4m}y^s(c\tau+d)^{-1/2}e^{2\pi i\left(\frac{a}{c}\frac{r^2}{4m}-\frac{sr}{2mc}+\frac{d}{c}\frac{s^2}{4m}\right)}\int_{0}^1 e^{-2\pi i\frac{mc}{c\tau+d}\left(x-\frac{r}{2mc}+s\frac{c\tau+d}{2mc}\right)^2}dx\nonumber\\ \label{theta rewrite 2}
&=&\sum_{\sigma\text{ mod }2m}\sum_{n\in\mathbb{Z}}\sum_{r=\rho\,\text{mod}(2m)} q^{(\sigma +2mn)^2/4m}y^{(\sigma+2mn)} e^{2\pi i\left(\frac{a}{c}\frac{r^2}{4m}-\frac{(\sigma+2mn)r}{2mc}+\frac{d}{c}\frac{(\sigma+2mn)^2}{4m}\right)}\nonumber\\
 && \times(c\tau+d)^{-1/2}\int_{0}^1 e^{-2\pi i\frac{mc}{c\tau+d}\left(x-\frac{r}{2mc}+(\sigma+2mn)\frac{c\tau+d}{2mc}\right)^2}dx.
\end{eqnarray}
 Using the decomposition $r=\rho+2m\alpha+2m l c$ with $0\leq\alpha\leq c-1$ and $l\in\mathbb{Z}$, we obtain
\begin{eqnarray}
 &&\sum_{r=\rho\,\text{mod}(2m)}e^{2\pi i\left(\frac{a}{c}\frac{r^2}{4m}-\frac{s r}{2mc}+\frac{d}{c}\frac{s^2}{4m}\right)} \int_{0}^1 e^{-2\pi i\frac{mc}{c\tau+d}\left(x-\frac{r}{2mc}-n\frac{d}{c}+(\sigma+2mn)\frac{c\tau+d}{2mc}\right)^2}dx\nonumber\\
&&=\sum_{\alpha=0}^{c-1}e^{2\pi i\left(\frac{a}{c}\frac{(\rho+2m\alpha)^2}{4m}-\frac{s (\rho+2m\alpha)}{2mc}+\frac{d}{c}\frac{s^2}{4m}\right)}\int_{-\infty}^{\infty} e^{-2\pi i\frac{mc}{c\tau+d}\left(x-\frac{\rho+2m\alpha}{2mc}-n\frac{d}{c}+(\sigma+2mn)\frac{c\tau+d}{2mc}\right)^2}dx\nonumber\\ \label{integral x}
&&=(c\tau+d)^{1/2}\frac{1}{(2mc i)^{1/2}}\sum_{\alpha=0}^{c-1}e^{2\pi i\left(\frac{a}{c}\frac{(\rho+2m\alpha)^2}{4m}-\frac{s (\rho+2m\alpha)}{2mc}+\frac{d}{c}\frac{s^2}{4m}\right)},
\end{eqnarray}
with $s=\sigma+2mn$, and we have used the fact that the sum over $r$ splits into a sum, first, over $\alpha$ and then over $l$, with the later being absorbed in a redefinition of $x$, extending the integral to the real line. On the other hand, we can show that the sum
\begin{equation}\label{appendix r goes to sigma}
\sum_{\alpha=0}^{c-1}\exp{\left[2\pi i\left(\frac{a}{c}\frac{(\rho+2m\alpha)^2}{4m}-\frac{s(\rho+2m\alpha)}{2mc}+\frac{d}{c}\frac{s^2}{4m}\right)\right]},
\end{equation}
with $s=\sigma+2mn$, is invariant under either $\rho\rightarrow \rho+2m l_1$ or $s\rightarrow s+2ml_2$, with $l_1,l_2\in \mathbb{Z}$. To see this, note that a shift of $\rho$ by $2ml$ in (\ref{appendix r goes to sigma}) is equivalent to a shift of $\alpha$ by $l$. Since we are summing over the equivalence class $\alpha\in \mathbb{Z}/c\mathbb{Z}$ this shift is innocuous in the sum over $\alpha$. To show invariance under $s\rightarrow s+2ml$, the first step is to rewrite the exponential in (\ref{appendix r goes to sigma}) as
\begin{eqnarray}
&&2\pi i\left(\frac{a}{c}\frac{(\rho+2m\alpha)^2}{4m}-\frac{s (\rho+2m\alpha)}{2mc}+\frac{d}{c}\frac{s^2}{4m}\right)=\nonumber\\ \label{sigma equiv class}
&&2\pi i\left(\frac{a}{c}\frac{(\rho+2m\alpha-d s)^2}{4m}+\frac{b\sigma (\rho+2m\alpha)}{2m}-\frac{bd}{4m}\sigma^2\right).
\end{eqnarray}
The shift of $s$ by $2ml$ can be absorbed in $\alpha$, shifting by $-d l$. Since the sum over $\alpha$ is independent of these shifts, as explained previously, we find that the sum (\ref{appendix r goes to sigma}) only depends on the equivalence class of $s$, that is, on $\sigma$.

Introducing the result (\ref{integral x}) back in (\ref{theta rewrite 2}) we obtain finally
\begin{eqnarray}
 (\theta_{m,\rho}|\gamma)(\tau,z)&=&\sum_{\sigma\text{ mod }2m}\sum_{n\in\mathbb{Z}}\sum_{r=\rho\,\text{mod}(2m)} q^{(\sigma +2mn)^2/4m}y^{(\sigma+2mn)} \frac{1}{(2mc i)^{1/2}}\sum_{\alpha=0}^{c-1}e^{2\pi i\left(\frac{a}{c}\frac{(\rho+2m\alpha)^2}{4m}-\frac{\sigma (\rho+2m\alpha)}{2mc}+\frac{d}{c}\frac{\sigma^2}{4m}\right)}\nonumber\\
&=&\sum_{\sigma\text{ mod }2m}\frac{1}{(2mc i)^{1/2}}\sum_{\alpha=0}^{c-1}e^{2\pi i\left(\frac{a}{c}\frac{(\rho+2m\alpha)^2}{4m}-\frac{\sigma (\rho+2m\alpha)}{2mc}+\frac{d}{c}\frac{\sigma^2}{4m}\right)}\theta_{m,\sigma}(\tau,z),
\end{eqnarray}
and hence by (\ref{def Kloos}) we conclude
\begin{equation}\label{Kloos rep}
 K_{\rho,\sigma}(\gamma)=\frac{1}{(2mc i)^{1/2}}\sum_{\alpha=0}^{c-1}e^{2\pi i\left(\frac{a}{c}\frac{(\rho+2m\alpha)^2}{4m}-\frac{\sigma (\rho+2m\alpha)}{2mc}+\frac{d}{c}\frac{\sigma^2}{4m}\right)},\;\gamma=\left(\begin{array}{cc}
a & b\\
c & d 
 \end{array}\right)\in SL(2,\mathbb{Z}).
\end{equation}

\subsection{Some properties of $K(\gamma)_{\rho\sigma}$}

In the following, we show explicitly that the the Kloosterman formula (\ref{Kloos rep}) obeys the group property
\begin{equation}\label{group property}
\sum_{\lambda=1}^{2m}K_{\rho,\lambda}(\gamma)K_{\lambda, \sigma}(\gamma')=K_{\rho,\sigma}(\gamma\gamma'),\qquad \gamma,\gamma'\in SL(2,\mathbb{Z}),
\end{equation}
as expected from the definition (\ref{def Kloos}). 

 To do that, we need a few other properties of the formula (\ref{Kloos rep}). From the previous analysis of (\ref{appendix r goes to sigma}), we can conclude 
\begin{equation}\label{prop1}
K_{\rho,\sigma}(\gamma)=K_{\rho+2m l,\sigma}(\gamma),\quad K_{\rho,\sigma}(\gamma)=K_{\rho,\sigma+2ml}(\gamma),\;l\in \mathbb{Z}.
\end{equation}

That is, the Kloosterman sums $K_{\rho,\sigma}$ (\ref{Kloos rep}) only depend on the equivalence class of $\rho\in \mathbb{Z}/2m\mathbb{Z}$ and $\sigma\in \mathbb{Z}/2m\mathbb{Z}$. 
 
 Given (\ref{prop1}), we can rewrite
\begin{equation}\label{rewriting 1}
\sum_{\lambda=1}^{2m}K_{\rho,\lambda}(\gamma)K_{\lambda, \sigma}(\gamma')=\frac{1}{cc'}\sum_{\lambda=1}^{2mcc'}K_{\rho,\lambda}(\gamma)K_{\lambda, \sigma}(\gamma'),
\end{equation}
with
\begin{equation}
\gamma=\left(\begin{array}{cc}
a & b\\
c & d
\end{array}\right),\; \gamma'=\left(\begin{array}{cc}
a' & b'\\
c' & d'
\end{array}\right)\in SL(2,\mathbb{Z})\nonumber.
\end{equation}
Then we have 
\begin{eqnarray}
&&\sum_{\lambda=1}^{2m}K_{\rho,\lambda}(\gamma)K_{\lambda, \sigma}(\gamma')=\nonumber\\
&&=\frac{1}{cc'}\frac{1}{2mi\sqrt{cc'}}\sum_{\alpha=0}^{c-1}\sum_{\beta=0}^{c'-1}\exp{\left[\pi i\frac{a}{2mc}(\rho+2m\alpha)^2+2\pi i m \frac{a'}{c'}\beta^2-2\pi i\frac{\sigma\beta}{c'}+\frac{\pi i}{2m}\frac{d'}{c'}\sigma^2\right] }\nonumber\\ \label{sum lambda}
&&\times\sum_{\lambda=1}^{2mcc'}\exp{\left[\pi i\frac{c''}{2mcc'}\lambda^2+\pi i\lambda\left(2\frac{a'}{c'}\beta-\frac{1}{mc}(\rho+2m\alpha)-\frac{\sigma}{mc'}\right)\right]},
\end{eqnarray}
with
\begin{equation}
\gamma\gamma'=\left(\begin{array}{cc}
a'' & b''\\
c'' & d''
\end{array}\right)\in SL(2,\mathbb{Z})\nonumber.
\end{equation}
In the sum (\ref{sum lambda}) we can use Gauss's reciprocity formula
\begin{equation}
\sum_{\lambda\text{ mod}(n)}\exp{\left[\pi i\frac{m}{n}\lambda^2+2\pi i \psi \lambda\right]}=\sqrt{\frac{i n}{m}}\sum_{\lambda\text{ mod}(m)}\exp{\left[-\pi i\frac{n}{m}(\lambda+ \psi)^2\right]},
\end{equation}
with $n,m\in \mathbb{Z}$, $nm\in 2\mathbb{Z}$ and $n\psi \in \mathbb{Z}$. Note that the sum over $\lambda$ in (\ref{sum lambda}) only depends on $\lambda\,\text{mod}(2mcc')$ and hence, using the reciprocity formula we can write
\begin{eqnarray}
&&\sum_{\lambda=1}^{2mcc'}\exp{\left[\pi i\frac{c''}{2mcc'}\lambda^2+\pi i\lambda\left(2\frac{a'}{c'}\beta-\frac{1}{mc}(\rho+2m\alpha)-\frac{\sigma}{mc'}\right)\right]}=\nonumber\\
&&\sqrt{\frac{2imcc'}{c''}}\sum_{\lambda=0}^{c''-1}\exp{\left[-2\pi i\frac{mcc'}{c''}\left(\lambda+\frac{a'}{c'}\beta-\frac{1}{2mc}(\rho+2m\alpha)-\frac{\sigma}{2mc'}\right)^2\right]}\nonumber.
\end{eqnarray}
Thus we find
\begin{eqnarray}
&&\sum_{\lambda=1}^{2m}K_{\rho,\lambda}(\gamma)K_{\lambda, \sigma}(\gamma')=\nonumber\\
&&=\frac{1}{cc'}\frac{1}{\sqrt{2m i c''}}\sum_{\alpha=0}^{c-1}\sum_{\beta=0}^{c'-1}\exp{\left[\pi i\frac{a}{2mc}(\rho+2m\alpha)^2+2\pi i m \frac{a'}{c'}\beta^2-2\pi i\frac{\sigma\beta}{c'}+\frac{\pi i}{2m}\frac{d'}{c'}\sigma^2\right] }\nonumber\\
&&\times \sum_{\lambda=0}^{c''-1}\exp{\left[-2\pi i\frac{mcc'}{c''}\left(\lambda+\frac{a'}{c'}\beta-\frac{1}{2mc}(\rho+2m\alpha)-\frac{\sigma}{2mc'}\right)^2\right]}.
\end{eqnarray}
Summing over $\lambda$ is the same as summing over equivalence classes of $\alpha$ since $\lambda-\alpha/c=(c\lambda-\alpha)/c$. Therefore, we can absorb the sum over $\lambda$ in $\alpha$ to obtain
\begin{eqnarray}
&&\sum_{\lambda=1}^{2m}K_{\rho,\lambda}(\gamma)K_{\lambda, \sigma}(\gamma')=\nonumber\\
&&=\frac{1}{cc'}\frac{1}{\sqrt{2m i c''}}\sum_{\alpha=0}^{cc''-1}\sum_{\beta=0}^{c'-1}\exp{\left[\pi i\frac{a}{2mc}(\rho+2m\alpha)^2+2\pi i m \frac{a'}{c'}\beta^2-2\pi i\frac{\sigma\beta}{c'}+\frac{\pi i}{2m}\frac{d'}{c'}\sigma^2\right] }\nonumber\\
&&\times\exp{\left[-2\pi i\frac{mcc'}{c''}\left(\frac{a'}{c'}\beta-\frac{1}{2mc}(\rho+2m\alpha)-\frac{\sigma}{2mc'}\right)^2\right]}\nonumber.
\end{eqnarray}
Similarly, since $\beta a'/c'-\sigma/2mc'=(2m a'\beta-\sigma)/2mc'$, the sum over $\beta$ can be traded by a sum over equivalence classes of $\sigma$. Moreover, under $\sigma \rightarrow \sigma +2m a'\beta$ we have 
\begin{equation}
2\pi i m \frac{a'}{c'}\beta^2-2\pi i\frac{\sigma\beta}{c'}+\frac{\pi i}{2m}\frac{d'}{c'}\sigma^2\rightarrow \frac{\pi i}{2m}\frac{d'}{c'}\sigma^2+2\pi i \mathbb{Z}.
\end{equation}
Since a shift of $\sigma$ by $2ml$ with $l\in \mathbb{Z}$ is innocuous in $K_{\lambda\sigma}(\gamma)$, the sum over $\beta$ gives an exact $c'$ factor. Therefore we get
\begin{eqnarray}
&&\sum_{\lambda=1}^{2m}K_{\rho,\lambda}(\gamma)K_{\lambda, \sigma}(\gamma')=\nonumber\\
&&=\frac{1}{c}\frac{1}{\sqrt{2m i c''}}\sum_{\alpha=0}^{cc''-1}\exp{\left[\pi i\frac{a}{2mc}(\rho+2m\alpha)^2+\frac{\pi i}{2m}\frac{d'}{c'}\sigma^2\right] }\nonumber\\
&&\times\exp{\left[-2\pi i\frac{mcc'}{c''}\left(\frac{1}{2mc}(\rho+2m\alpha)+\frac{\sigma}{2mc'}\right)^2\right]}\nonumber\\
\label{sum alpha}&&=\frac{1}{c}\frac{1}{\sqrt{2m i c''}}\sum_{\alpha=0}^{cc''-1}\exp{\left[2\pi i\left(\frac{a''}{c''}\frac{(\rho+2m\alpha)^2}{4m}-\frac{\sigma (\rho+2m\alpha)}{2mc''}+\frac{d''}{c''}\frac{\sigma^2}{4m}\right)\right]}\nonumber.\\
{}
\end{eqnarray}
We now use the property that the sum over $\alpha$ in (\ref{sum alpha}) depends only on the equivalence class $\alpha \,\text{mod}(c'')$.  This gives a factor of $c$. We obtain finally
\begin{eqnarray}
&&\sum_{\lambda=1}^{2m}K_{\rho,\lambda}(\gamma)K_{\lambda, \sigma}(\gamma')=\nonumber\\
&&=\frac{1}{\sqrt{2m i c''}}\sum_{\alpha=0}^{c''-1}\exp{\left[2\pi i\left(\frac{a''}{c''}\frac{(\rho+2m\alpha)^2}{4m}-\frac{\sigma (\rho+2m\alpha)}{2mc''}+\frac{d''}{c''}\frac{\sigma^2}{4m}\right)\right]}=K_{\rho\sigma}(\gamma\gamma'),\nonumber\\
{}
\end{eqnarray}
as we wanted to show.
\end{appendices}

\bibliographystyle{JHEP}
\bibliography{measure2}

\providecommand{\href}[2]{#2}\begingroup\raggedright\begin{thebibliography}{10}

\bibitem{Kloos}
H.~Kloosterman, {\it {On the representation of numbers in the form
  {$ax^2+by^2+cw^2+dz^2$}}},  {\em Acta Math.} {\bf 49} (1926) 407--464.

\bibitem{Rademacher-1938}
H.~Rademacher, {\it The fourier coefficients of the modular invariant
  {$J(\tau)$}},  {\em American Journal of Mathematics} {\bf 60} (1938), no.~2
  501--512.

\bibitem{Zuckerman-Rademacher}
H.~Rademacher and H.~S. Zuckerman, {\it On the fourier coefficients of certain
  modular forms of positive dimension},  {\em Annals of Mathematics} {\bf 39}
  (1938), no.~2 433--462.

\bibitem{Dabholkar:2014ema}
A.~Dabholkar, J.~Gomes, and S.~Murthy, {\it {Nonperturbative black hole entropy
  and Kloosterman sums}},  {\em JHEP} {\bf 1503} (2015) 074,
  [\href{http://xxx.lanl.gov/abs/1404.0033}{{\tt arXiv:1404.0033}}].

\bibitem{Sen:2008vm}
A.~Sen, {\it {Quantum Entropy Function from AdS(2)/CFT(1) Correspondence}},
  \href{http://xxx.lanl.gov/abs/0809.3304}{{\tt arXiv:0809.3304}}.

\bibitem{Dabholkar:2010uh}
A.~Dabholkar, J.~Gomes, and S.~Murthy, {\it {Quantum black holes, localization
  and the topological string}},  \href{http://xxx.lanl.gov/abs/1012.0265}{{\tt
  arXiv:1012.0265}}.

\bibitem{Dabholkar:2011ec}
A.~Dabholkar, J.~Gomes, and S.~Murthy, {\it {Localization \&; Exact
  Holography}},  {\em JHEP} {\bf 1304} (2013) 062,
  [\href{http://xxx.lanl.gov/abs/1111.1161}{{\tt arXiv:1111.1161}}].

\bibitem{Gupta:2012cy}
R.~K. Gupta and S.~Murthy, {\it {All solutions of the localization equations
  for N=2 quantum black hole entropy}},  {\em JHEP} {\bf 1302} (2013) 141,
  [\href{http://xxx.lanl.gov/abs/1208.6221}{{\tt arXiv:1208.6221}}].

\bibitem{Dijkgraaf:2000fq}
R.~Dijkgraaf, J.~M. Maldacena, G.~W. Moore, and E.~P. Verlinde, {\it {A Black
  hole Farey tail}},  \href{http://xxx.lanl.gov/abs/hep-th/0005003}{{\tt
  hep-th/0005003}}.

\bibitem{Manschot:2007ha}
J.~Manschot and G.~W. Moore, {\it {A Modern Farey Tail}},  {\em
  Commun.Num.Theor.Phys.} {\bf 4} (2010) 103--159,
  [\href{http://xxx.lanl.gov/abs/0712.0573}{{\tt arXiv:0712.0573}}].

\bibitem{Eichler:1985ja}
M.~Eichler and D.~Zagier, {\em {The Theory of Jacobi Forms}}.
\newblock Birkh{\"a}user, 1985.

\bibitem{Dabholkar:2012nd}
A.~Dabholkar, S.~Murthy, and D.~Zagier, {\it {Quantum Black Holes, Wall
  Crossing, and Mock Modular Forms}},
  \href{http://xxx.lanl.gov/abs/1208.4074}{{\tt arXiv:1208.4074}}.

\bibitem{Ferrari:2017msn}
F.~Ferrari and V.~Reys, {\it {Mixed Rademacher and BPS Black Holes}},
  \href{http://xxx.lanl.gov/abs/1702.0275}{{\tt arXiv:1702.0275}}.

\bibitem{Maldacena:1999bp}
J.~M. Maldacena, G.~W. Moore, and A.~Strominger, {\it {Counting BPS black holes
  in toroidal Type II string theory}},
  \href{http://xxx.lanl.gov/abs/hep-th/9903163}{{\tt hep-th/9903163}}.

\bibitem{Lisa92}
L.~C. Jeffrey, {\it {Chern-Simons-Witten invariants of lens spaces and torus
  bundles, and the semiclassical approximation}},  {\em Commun. Math. Phys.}
  {\bf 147} (1992) 563--604.

\bibitem{Witten:1988hf}
E.~Witten, {\it {Quantum Field Theory and the Jones Polynomial}},  {\em
  Commun.Math.Phys.} {\bf 121} (1989) 351.

\bibitem{10.2307/1969082}
H.~D. Kloosterman, {\it The behavior of general theta functions under the
  modular group and the characters of binary modular congruence groups. i},
  {\em Annals of Mathematics} {\bf 47} (1946), no.~3 317--375.

\bibitem{Gomes:2017eac}
J.~Gomes, {\it {Quantum Black Hole Entropy, Localization and the Stringy
  Exclusion Principle}},  \href{http://xxx.lanl.gov/abs/1705.0195}{{\tt
  arXiv:1705.0195}}.

\bibitem{Maldacena:1998bw}
J.~M. Maldacena and A.~Strominger, {\it {AdS(3) black holes and a stringy
  exclusion principle}},  {\em JHEP} {\bf 12} (1998) 005,
  [\href{http://xxx.lanl.gov/abs/hep-th/9804085}{{\tt hep-th/9804085}}].

\bibitem{Gaiotto:2006ns}
D.~Gaiotto, A.~Strominger, and X.~Yin, {\it {From AdS(3)/CFT(2) to black holes
  / topological strings}},  {\em JHEP} {\bf 09} (2007) 050,
  [\href{http://xxx.lanl.gov/abs/hep-th/0602046}{{\tt hep-th/0602046}}].

\bibitem{Gaiotto:2006wm}
D.~Gaiotto, A.~Strominger, and X.~Yin, {\it {The M5-brane elliptic genus:
  Modularity and BPS states}},  {\em JHEP} {\bf 08} (2007) 070,
  [\href{http://xxx.lanl.gov/abs/hep-th/0607010}{{\tt hep-th/0607010}}].

\bibitem{Gomes:2015xcf}
J.~Gomes, {\it {Exact Holography and Black Hole Entropy in N=8 and N=4 String
  Theory}},  \href{http://xxx.lanl.gov/abs/1511.0706}{{\tt arXiv:1511.0706}}.

\bibitem{Murthy:2009dq}
S.~Murthy and B.~Pioline, {\it {A Farey tale for N=4 dyons}},  {\em JHEP} {\bf
  09} (2009) 022, [\href{http://xxx.lanl.gov/abs/0904.4253}{{\tt
  arXiv:0904.4253}}].

\bibitem{kirk1990}
P.~Kirk and E.~Klassen, {\it {Chern-Simons invariants of 3-manifolds and
  representation spaces of knot groups}},  {\em Mathematische Annalen} {\bf
  287} (1990), no.~1 343--367.

\bibitem{Elitzur:1989nr}
S.~Elitzur, G.~W. Moore, A.~Schwimmer, and N.~Seiberg, {\it {Remarks on the
  Canonical Quantization of the Chern-Simons-Witten Theory}},  {\em Nucl.Phys.}
  {\bf B326} (1989) 108.

\bibitem{Hansen:2006wu}
J.~Hansen and P.~Kraus, {\it {Generating charge from diffeomorphisms}},  {\em
  JHEP} {\bf 0612} (2006) 009,
  [\href{http://xxx.lanl.gov/abs/hep-th/0606230}{{\tt hep-th/0606230}}].

\bibitem{Maldacena:1997de}
J.~M. Maldacena, A.~Strominger, and E.~Witten, {\it {Black hole entropy in
  M-theory}},  {\em JHEP} {\bf 12} (1997) 002,
  [\href{http://xxx.lanl.gov/abs/hep-th/9711053}{{\tt hep-th/9711053}}].

\bibitem{Sen:2008sp}
A.~Sen, {\it {U-duality Invariant Dyon Spectrum in type II on T**6}},  {\em
  JHEP} {\bf 0808} (2008) 037, [\href{http://xxx.lanl.gov/abs/0804.0651}{{\tt
  arXiv:0804.0651}}].

\bibitem{Banerjee:2008pu}
S.~Banerjee, A.~Sen, and Y.~K. Srivastava, {\it {Partition Functions of Torsion
  $ > 1$ Dyons in Heterotic String Theory on {$T^6$}}},
  \href{http://xxx.lanl.gov/abs/0802.1556}{{\tt 0802.1556}}.

\bibitem{Dabholkar:2008zy}
A.~Dabholkar, J.~Gomes, and S.~Murthy, {\it {Counting all dyons in N =4 string
  theory}},  \href{http://xxx.lanl.gov/abs/0803.2692}{{\tt arXiv:0803.2692}}.

\bibitem{Sen:2009vz}
A.~Sen, {\it {Arithmetic of Quantum Entropy Function}},  {\em JHEP} {\bf 0908}
  (2009) 068, [\href{http://xxx.lanl.gov/abs/0903.1477}{{\tt
  arXiv:0903.1477}}].

\bibitem{Sen:2009gy}
A.~Sen, {\it {Arithmetic of N=8 Black Holes}},  {\em JHEP} {\bf 02} (2010) 090,
  [\href{http://xxx.lanl.gov/abs/0908.0039}{{\tt arXiv:0908.0039}}].

\bibitem{Zagier1989}
S.~N.-P. Zagier, Don, {\it A trace formula for jacobi forms.},  {\em Journal
  für die reine und angewandte Mathematik} {\bf 393} (1989) 168--198.

\end{thebibliography}\endgroup

\end{document}